\documentclass[aps,pra,showpacs,twocolumn]{revtex4-1}
\usepackage{graphicx,epstopdf}
\usepackage{multirow}
\usepackage{mathtools}
\usepackage{amsmath}
\usepackage{amssymb,mathptmx}
\usepackage{fouridx}
\usepackage{graphicx}
\usepackage{amsthm}
\usepackage{eucal}
\usepackage{bm}
\usepackage{upgreek}
\usepackage{framed}
\usepackage{mathrsfs}
\usepackage{latexsym}
\usepackage{float}
\usepackage{subfigure}
\usepackage{color}
\usepackage{graphicx} 
\usepackage{bm,hyperref,tikz}
\usepackage[utf8]{inputenc} 
\usepackage{bbold}
\usepackage{bbm}

\newcommand{\be}{\begin{equation}}
	\newcommand{\ee}{\end{equation}}
\newcommand{\bea}{\begin{eqnarray}}
	\newcommand{\eea}{\end{eqnarray}}
\newcommand{\ba}{\begin{array}}
	\newcommand{\ea}{\end{array}}

\newcommand{\bla}{\color{black}}
\newcommand{\ket}[1]{\left\vert#1\right\rangle}

\newcommand{\Ba}{\mathrm{B}_\alpha\,}
\newcommand{\BI}{\mathrm{B}_\mathrm{I}}

\newcommand{\E}[1]{\mathrm{e}^{\mbox{\footnotesize$#1$}}}

\begin{document}	
	
\title{Resource-efficient topological fault-tolerant  quantum computation with hybrid entanglement of light}
	
\author{S. Omkar}
\email{omkar.shrm@gmail.com}
\affiliation{Department of Physics and Astronomy, Seoul National University, 08826 Seoul, Republic of Korea}

\author{Yong Siah Teo}
\affiliation{Department of Physics and Astronomy, Seoul National University, 08826 Seoul, Republic of Korea}

\author{Hyunseok Jeong}
\email{h.jeong37@gmail.com}
\affiliation{Department of Physics and Astronomy, Seoul National University, 08826 Seoul, Republic of Korea}

\begin{abstract}

We propose an all-linear-optical scheme to ballistically generate a cluster state for measurement-based topological fault-tolerant  quantum computation using hybrid photonic qubits entangled in a continuous-discrete domain. Availability of near-deterministic Bell-state measurements on hybrid qubits is exploited for the purpose. In the presence of photon losses, we show that our scheme leads to a significant enhancement in both  tolerable photon-loss rate and resource overheads. More specifically, we report a photon-loss threshold of $\sim3.3\times 10^{-3}$, which is higher than those of known optical schemes under a reasonable error model.
Furthermore, resource overheads to achieve logical error rate of $10^{-6} (10^{-15})$ is estimated to be $\sim8.5\times10^{5} (1.7\times10^{7})$ which is significantly less by multiple orders of magnitude compared to other reported values in the literature.

\end{abstract}
	
\maketitle

Errors during quantum information processing are unavoidable, and they are  a major obstacle against practical implementations of quantum computation (QC)~\cite{NC10}. Quantum error correction (QEC)~\cite{LD13} permits scalable QC with faulty qubits and gates provided the noise is below a certain threshold. The noise threshold is determined by the details of the  implementing  scheme and the noise model.

Measurement-based topological fault-tolerant (FT)  QC~\cite{RHG06} on a cluster state
provides a high error threshold of $0.75\%$ against computational errors~\cite{ RHG07, RH07}. Additionally, it can tolerate qubit losses~\cite{BS10, WF14} and missing edges~\cite{LDSB10}; thus, it would be suitable for practical large-scale QC. However, there is a trade-off  between the tolerable computational error rate, and the tolerable level of qubit losses and missing edges. A cluster state $\ket{\mathcal{C}}$, over a collection of qubits $\mathcal{C}$, is the state stabilized by  operators  $X_a\bigotimes_{b\in \rm{ nh}(a)} Z_b$, where $a,b\in\mathcal{C}$, $Z_i$ and $X_i$ are the Pauli operators on the $i$th qubit, and nh(a) represents the adjacent neighborhood of qubit $a\in\mathcal{C}$ \cite{BR01}. It has the form:  $\ket{\mathcal{C}}=\prod_{b\in \rm{ nh}(a)}\textrm{CZ}_{a,b}\ket{+}_a\ket{+}_b,~\forall a\in \mathcal{C}$, where CZ is the controlled-Z gate,
 $\ket{\pm}=(\ket{0}\pm\ket{1})/\sqrt{2}$, 
and $\{\ket{0},\ket{1}\}$ are eigenstates of $Z$.
Here, we consider the Raussendorf cluster state  $\ket{\mathcal{C}_\mathcal{L}}$ \cite{RHG06} on a cubic lattice $\mathcal{L}$ with qubits  mounted on its faces and edges.

The linear optical platform has the advantage of quick gate operations compared to their decoherence time~\cite{RP10}. Unfortunately, schemes based on discrete variables (DV) like photon polarizations  suffer from the drawback that the entangling operations (EOs), typically implemented by Bell-state measurements, are  nondeterministic~\cite{BR05}.
This leaves the edges corresponding to all failed EOs missing, and beyond a certain failure rate the cluster state cannot support QC. References~\cite{Nielsen04, DHN06, LDSB10, FT10, HFJR10} tackle this shortcoming with a repeat-until-success strategy. However, this strategy incurs heavy resource overheads  in terms of both qubits and EO trials, and the overheads grow exponentially as the success rate of EO falls~\cite{LDSB10}. Moreover, conditioned on the outcome of the EO, all other redundant qubits must be removed {\it via} measurements~\cite{FT10} which would add to undesirable resource overheads. These schemes also require  active switching to select successful outcomes of EOs and feed them to the next stage, which is known to have an adverse effect on the photon-loss threshold for FTQC~\cite{LHMB15}.
DV-based optical EOs have a success rate of 50\% that can be  further  boosted with additional resources like single photons~\cite{EL14}, Bell states~\cite{G11} and the squeezing operation~\cite{ZL13}. 
Reference~\cite{GSBR15} uses EOs with boosted success rate of 75\% to build cluster states.  
This can be further enhanced by allotting more resources.
Coherent-state qubits, composed of coherent states  $|\pm\alpha\rangle$ of amplitudes $\pm\alpha$,
 enable one to perform nearly deterministic Bell-state measurements and universal QC using linear optics~\cite{JK02, RGM+03},
while this approach is generally more vulnerable to losses~\cite{LRH08, RP10}.
Along this line, a scheme to generate cluster states for topological QC was suggested, but the value of $\alpha$ required to build a cluster state of sufficiently high fidelity is unrealistically large as $\alpha>20$~\cite{MR11}. A hybrid qubit using both DV and continuous-variable (CV) states of light, i.e., polarized single photons and coherent states was introduced to take advantages of both the approaches~\cite{LJ13}.

We propose an all-linear-optical measurement-based FT hybrid topological QC (HTQC) scheme on  $\ket{\mathcal{C}_\mathcal{L}}$ of hybrid qubits.
The logical basis for a hybrid qubit is defined as $\{\ket{\alpha}\ket{\textsc{h}}\equiv\ket{0_{\rm L}},\ket{-\alpha}\ket{\textsc{v}}\equiv\ket{1_{\rm L}}\}$,
where $\ket{\textsc{h}}$ and $\ket{\textsc{v}}$ are single-photon states with  horizontal and vertical polarizations in the $Z$ direction. The issues with indeterminism of EOs on DVs \cite{DHN06, LDSB10, FT10, HFJR10} and  poor fidelity of  the cluster states with CVs~\cite{MR11} are then overcome.  Crucial to our scheme is a near-deterministic hybrid Bell-state measurement (HBSM) on hybrid qubits  using two photon number parity detectors (PNPDs) and two  on-off photodetectors (PDs), which is distinct from the previous version that requires two additional PDs to complete a teleportation protocol  \cite{LJ13}. 
We only need HBSMs acting on three-hybrid-qubit cluster states to generate $\ket{\mathcal{C}_\mathcal{L}}$ without any active switching and  feed-forward. The outcomes of HBSMs are noted to interpret the measurement results during QEC and QC. In this sense, our scheme is {\it ballistic} in nature.
Both CV and DV modes of hybrid qubits support the HBSMs to build  $\ket{\mathcal{C}_\mathcal{L}}$, while only  DV modes suffice for QEC and QC. This means that {\it only} on-off PDs for DV modes are required once  $\ket{\mathcal{C}_\mathcal{L}}$ is generated. 
 In addition, 
 photon loss is ubiquitous~\cite{RP10} that  causes dephasing 
such as in~\cite{LJ13, LRH08, LPRJ15}.  We analyze the performance of our scheme against photon losses  and  compare it with the known all-optical schemes.

\begin{figure}[t]
\includegraphics[width=.8\columnwidth]{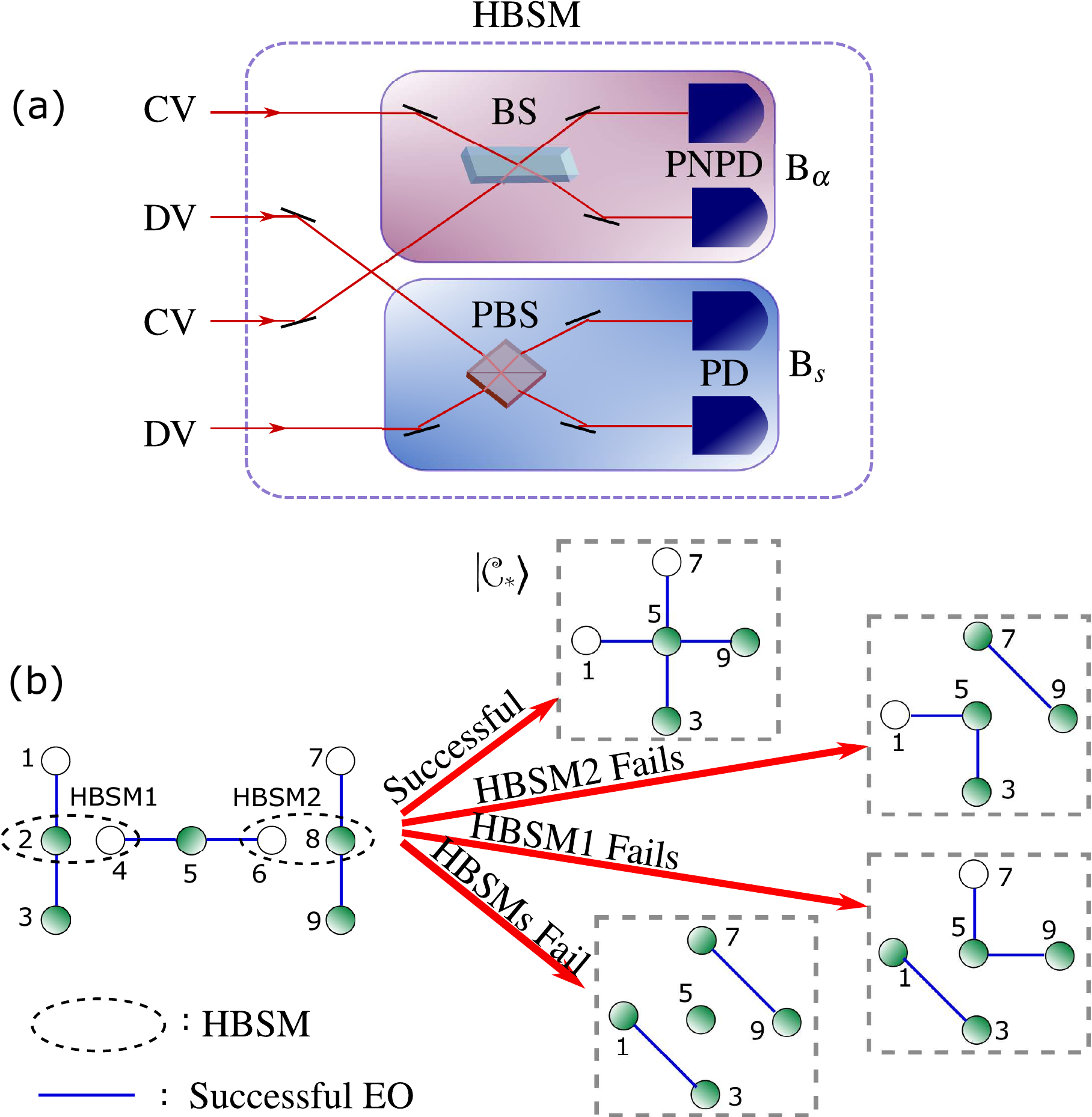}
\caption{(a) $\Ba$ acts on  CV modes and fails when neither of the two PNPDs click. The failure rate of a $\Ba$ on the hybrid qubits is $\E{-2\alpha^2}$. $\rm{B}_{\rm{s}}$ acts on  DV modes and is successful with probability $1/2$ only when both the PDs click.  (b) The three-hybrid-qubit cluster with one unfilled circle represents $\ket{\mathcal{C}_3}$, while that with two represents $\ket{\mathcal{C}_{3^\prime}}$ in Eq.~(\ref{eq:offline}). An unfilled circle means a difference by a Hadamard transform from the original three-qubit cluster (see Appendix~\ref{app:clus3}). 
 Success of both HBSMs creates a star cluster $\ket{\mathcal{C}_\ast}$ and other cases lead to distorted star clusters as shown. 
}
\label{fig:star}
\end{figure}

\begin{figure}[t]
	\includegraphics[width=.8\columnwidth]{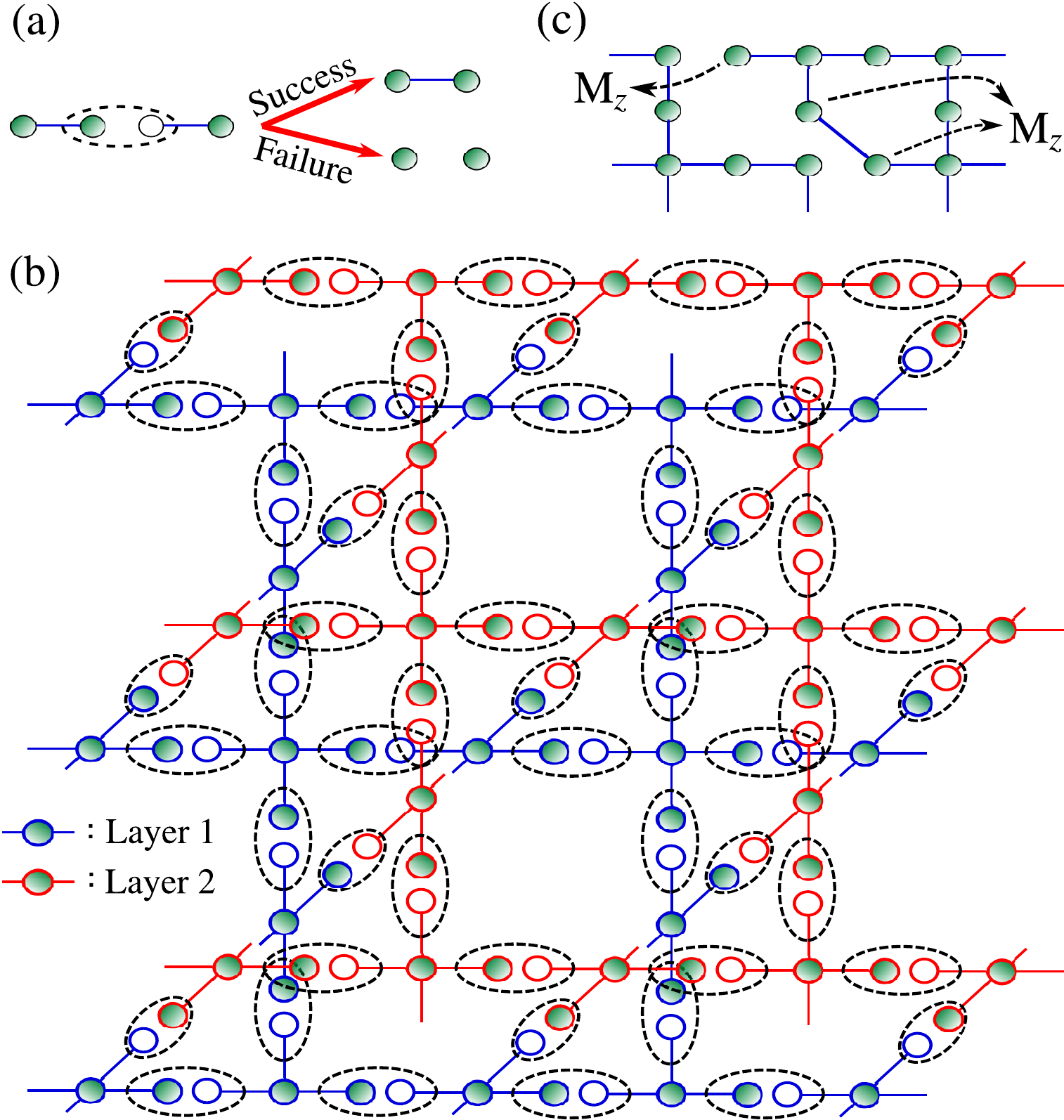}
	\caption{(a) When connecting $\ket{\mathcal{C}_\ast}$'s, a successful HBSM creates an edge between hybrid qubits whereas a failed HBSM leaves the edge missing.
(b) 3D  illustration of building two layers of $\ket{\mathcal{C}_\mathcal{L}}$ for practical HTQC with  $\ket{\mathcal{C}_\ast}$'s and HBSMs to connect them. (c) A  {\it diagonal edge} is created due to failure of an HBSM corresponding to $\ket{\mathcal{C}_\ast}$ and a missing edge  is due to failure of an HBSM while connecting them. A single layer of $\ket{\mathcal{C}_\mathcal{L}}$ is shown for convenience, and ${\rm  M_z}$ is measurement in Z-basis.
}
	\label{fig:2layer}
\end{figure}

\emph{Physical platform for $\ket{\mathcal{C}_\mathcal{L}}$.---}
To ballistically build a $\ket{\mathcal{C}_\mathcal{L}}$, we begin with 
hybrid qubits, in the form $(\ket{\textsc{h}}|\alpha\rangle+\ket{\textsc{v}}|-\alpha\rangle)/\sqrt{2}=(|0_{\rm L}\rangle+|1_{\rm L}\rangle)/\sqrt{2}\equiv|+_{\rm L}\rangle$, as raw resources of our scheme. In fact, this type of hybrid qubits and with slight variant forms (with the vacuum and single photon instead of $|H\rangle$ and $|V\rangle$) were generated in recent experiments~\cite{SUT+18, JZK+14, MHL+14}, which can also be used for QC in the same way as in~\cite{LJ13} even with higher  fidelities and success probabilities of teleportation \cite{KLJ16}.
A hybrid qubit can also be generated using  a Bell-type photon pair, a coherent-state superposition, linear optical elements and four PDs~\cite{KJ15}.

The HBSM introduced in this Letter consists of two types of measurements, $\Ba$  and $\rm B_s$, acting on CV and DV modes, respectively. 
A Bell-state measurement for coherent-state qubits \cite{JKL01}, $ \Ba$, comprises of a beam splitter (BS) and two PNPDs, whereas $\rm B_s$ has a polarizing BS (PBS) and  two PDs  as shown in Fig.~\ref{fig:star}(a). 
The  failure rate for an HBSM turns out to be $p_f=\E{-2\alpha^2}\!\!/2$ (see  Appendix~\ref{app:bsm} and also~\cite{LJ13}) that rapidly approaches zero with growing $\alpha$.
The first and only nondeterministic step of our protocol is to prepare two kinds of  three-hybrid-qubit cluster states,
\bea
\label{eq:offline}
\ket{\mathcal{C}_3}_{abc}&=&
\frac{1}{2}\big(\ket{0_{\rm L}}_a\ket{0_{\rm L}}_b\ket{0_{\rm L}}_c +\ket{0_{\rm L}}_a\ket{0_{\rm L}}_b\ket{1_{\rm L}}_c \nonumber\\
&&+\ket{1_{\rm L}}_a\ket{1_{\rm L}}_b\ket{0_{\rm L}}_c -\ket{1_{\rm L}}_a\ket{1_{\rm L}}_b\ket{1_{\rm L}}_c\big),\nonumber\\
\ket{\mathcal{C}_{3^\prime}}_{abc}&=&
\frac{1}{\sqrt{2}}\big(\ket{0_{\rm L}}_{a}\ket{0_{\rm L}}_{b}\ket{0_{\rm L}}_{c} + \ket{1_{\rm L}}_{a}\ket{1_{\rm L}}_{b}\ket{1_{\rm L}}_{c}\big)~
\eea
using four hybrid qubits, two $\rm B_\alpha$'s and a $\rm B_I$ as detailed in Appendix~\ref{app:clus3}  . (Here, $\rm B_I$ is a type-I fusion gate using two PBSs, two PDs and a $\pi/2$-rotator, of which the success probability is 1/2. See  Appendix~\ref{app:bsm}   for details.) As shown  in Fig.~\ref{fig:star}(b), an HBSM is performed on modes $2$ and $4$ of  $\ket{\mathcal{C}_{3}}_{123}$ and $\ket{\mathcal{C}_{3^\prime}}_{456}$, and the other HBSM is performed  similarly  between $\ket{\mathcal{C}_{3^\prime}}_{456}$ and $\ket{\mathcal{C}_{3}}_{789}$, which produces a \emph{star cluster}, $\ket{\mathcal{C}_\ast}$, with a high success probability. {\it Simultaneously}, the star clusters are connected  using HBSMs to form layers of $\ket{\mathcal{C}_\mathcal{L}}$ as depicted in Fig. \ref{fig:2layer}(b). 
As the third dimension of  $\ket{\mathcal{C}_\mathcal{L}}$ is time simulated, in practice only two physical layers suffice for QC~\cite{RHG07}.

Notably, different outcomes of HBSMs and failures during this process can be compensated during  QEC as explained below. 
As HBSMs have four possible outcomes from $\Ba$, the built cluster state is equivalent to $\ket{\mathcal{C}_\mathcal{L}}$ up to local Pauli operations. This can be compensated by  accordingly making   bit flips to the  measurement outcomes during QEC. This is achieved by classical processing and no additional quantum resources are required.
As shown in Fig.~\ref{fig:star}(b), failure(s) of HBSMs result(s) in a  {\it deformed} star cluster  with diagonal edge(s) instead of four proper edges stretching from the central qubit. The final cluster state $\ket{\mathcal{C}_\mathcal{L}}$ inherits these diagonal edges as shown in Fig.~\ref{fig:2layer}(c) with a {\it disturbed} stabilizer structure. 
However, failures of  HBSMs are  heralded, which reveals the locations of such diagonal edges.  These diagonal edges can be  removed by adaptively measuring the hybrid qubits in $Z$-basis (${\rm M}_Z$),  as shown in Fig.~\ref{fig:2layer}(c), restoring back the stabilizer structure of $\ket{\mathcal{C}_\mathcal{L}}$.
Failure of  HBSMs for connecting $\ket{\mathcal{C}_*}$'s simply leaves the edges missing as shown in Fig.~2(a) without distorting the stabilizer structure.

\emph{ Noise model.---} 
Let $\eta$ be the photon-loss rate due to imperfect sources and detectors, absorptive optical components and storages. In HTQC,  the effect of photon loss is threefold  (see Appendix~\ref{app:loss}    and  also~\cite{LJ13}) that (i) causes dephasing of hybrid qubits i.e., phase-flip errors $Z$, a form of computational error, with rate $p_Z=[1-(1-\eta)\,\E{-2\eta\alpha^2}]/2$, (ii) lowers the success rate of HBSM and (iii) makes hybrid qubits leak out of the logical basis.
Quantitatively, $p_f$ increases to $(1+\eta)\E{-2\alpha^{\prime 2}}/2$, where $\alpha^\prime=\sqrt{1-\eta}\alpha$. Thus, for a given $\eta$ and growing $\alpha$ we face a trade-off between the desirable success rate of HBSM and the detrimental dephasing rate $p_Z$.   

Further, like the type-II fusion gate in~\cite{VBR08},  ${\rm B_s}$  does not introduce computational errors during photon loss.  However, the action of $\Ba$ on the lossy hybrid qubits introduces additional dephasing as shown in  as shown in the   Appendix~\ref{app:ba}. 
To clarify, like DV schemes \cite{HFJR10}, photon loss does not imply hybrid-qubit loss. In many FTQC schemes $\eta$ has a typical operational value of $\sim10^{-3}$ (on the higher side)~\cite{LPRJ15, DHN06, HHGR10, Cho07}, i.e.,  $\eta\ll1$. 
The probability of  hybrid-qubit loss
due to photon loss, $\eta \E{-\alpha^{\prime 2}}$
(the overlap between a lossy hybrid qubit and the vacuum),
  is then very small compared to $p_f$ and  negligible to HTQC.  

\emph{Measurement-based HTQC.---}
Once the faulty
cluster state is built with missing and diagonal edges, and phase-flip errors on the constituent hybrid qubits, measurement-based HTQC is performed  by making sequential single-qubit measurements in  $X$ and $Z$ bases. 
A few chosen ones are measured in $Z$-basis to create  defects, 
and the rest are measured in X-basis for error syndromes during QEC and for effecting the  Clifford gates on the logical states of  $\ket{\mathcal{C}_\mathcal{L}}$. For Magic state distillation, measurements are made in  $(X\pm Y)/\sqrt{2}$ basis~\cite{ RHG06, RH07,RHG07}.
All these measurements are accomplished by measuring only  polarizations of  DV modes in the respective basis.    
These measurement outcomes should be interpreted with respect to the recorded HBSM outcomes as mentioned earlier.   

\emph{Simulations.---} 
Simulation of topological QEC is performed using AUTOTUNE \cite{FWMR12} (see  Appendix~\ref{app:sim} for a brief description). Only the central hybrid qubit of $\ket{\mathcal{C}_\ast}$ remains in the cluster and the rest are utilized by HBSMs. The $\ket{\mathcal{C}_\ast}$'s are arranged as shown in  Fig.~\ref{fig:2layer}.   Next, all  hybrid qubits  are subjected to dephasing of rate $p_Z$ following which EOs are performed using HBSMs. The action of $\Ba$ in HBSM dephases the adjacent remaining hybrid qubits, which can be modeled as  applying  $\{Z\otimes I,I\otimes Z\}$ with  rate $p_Z$. The technical details of the action of $\Ba$ under photon loss are presented in the Appendix~\ref{app:ba}. This concludes the simulation of building noisy $\ket{\mathcal{C}_\mathcal{L}}$. Further, the hybrid qubits waiting to undergo measurements as a part of QEC attract dephasing, and rate $p_Z$ again is assigned. During QEC, $X$-measurement outcomes used for syndrome extraction could be  erroneous. This error too is assigned  rate $p_Z$. Due to photon losses, the hybrid qubits leak out of the logical basis  failing the measurements on DV modes. This  leakage  is also assigned $p_Z$, which only overestimates $\eta$.

One missing edge due to failed HBSMs can be mapped to two missing hybrid qubits \cite{LDSB10}. Improving on this, by adaptively performing ${\rm M}_Z$ (Fig.~\ref{fig:2layer}(c)) on one of the hybrid qubits associated with a missing edge, this edge can be modeled with a missing qubit~\cite{AAG+18}.  
Then, QEC is carried out as in the case of missing qubits~\cite{BS10}.
In constructing  $\ket{\mathcal{C}_{\mathcal{L}}}$, an equal number of HBSMs are required for building $\ket{\mathcal{C}_\ast}$ and for connecting them. A failure of an HBSM  during the former process corresponds to  two hybrid-qubit losses, and the latter case to one (Fig.~\ref{fig:2layer}(c)).
Therefore, on average 1.5 hybrid qubits per  HBSM failure are lost.
 Percolation threshold for $\ket{\mathcal{C}_\mathcal{L}}$ is $0.249$ fraction of missing qubits~\cite{BS10, LZ98,PTEG19}, which corresponds to $\alpha\approx0.7425$ (when no computational error is tolerated, i.e., $\eta=0$), the critical value of $\alpha$ below which  HTQC becomes impossible. 

\begin{figure}
\includegraphics[width=1\linewidth]{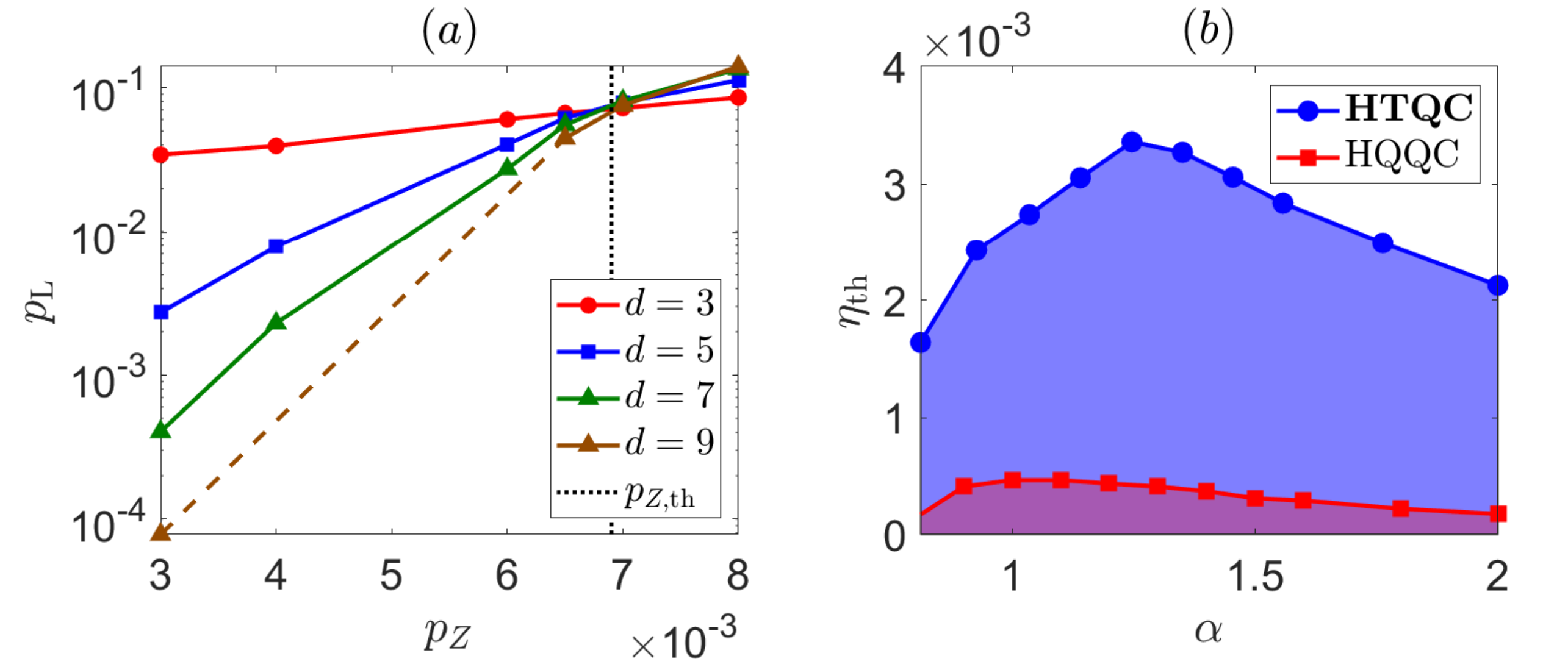}
\caption{(a) Logical error rate $p_{\rm L}$ is plotted against the dephasing rate $p_Z$ for coherent-state amplitude $\alpha_{\rm opt}=1.247$ 
 and code distances $d=3,5,7$. The intersecting point of these curves corresponds to the threshold dephasing rate $p_{Z,\rm th}$. (b) The tolerable photon-loss rate $\eta_{\rm{th}}$ is plotted against coherent-state amplitude $\alpha$. 
The behavior of the curve is due to the trade-off between the success rate of HBSM and dephasing rate $p_Z$ with growing $\alpha$. 
As we increase $\alpha$, both the success rate and $p_Z$ increase; but the former dominates and leads to an increase in $\eta_{\rm th}$.
When $\alpha>1.247$, $p_Z$  dominates and causes  $\eta_{\rm th}$ to decrease. Compared to the non-topoligical HQQC~\cite{LJ13}, HTQC has an order of higher value for $\eta_{\rm th}$.}
\label{fig:result}
\end{figure}

\emph{Results.---}
The logical error  rate $p_{\rm L}$ (failure rate of topological QEC~\cite{RHG07}) was determined against various values of  $p_Z$ for $\ket{\mathcal{C}_\mathcal{L}}$ of  code distances $d=3,5,7$.  This was repeated for various values of $p_f$ which correspond to different values of $\alpha$.  Figure~\ref{fig:result}(a) shows the simulation results for $\alpha_{\rm opt}=1.247$ in which the intersection point of the curves corresponds to the threshold dephasing rate $p_{Z,\rm th}$. The photon-loss threshold $\eta_{\rm th}$ is determined using the expression for $p_Z$.

\begin{figure}
\includegraphics[width=1\linewidth]{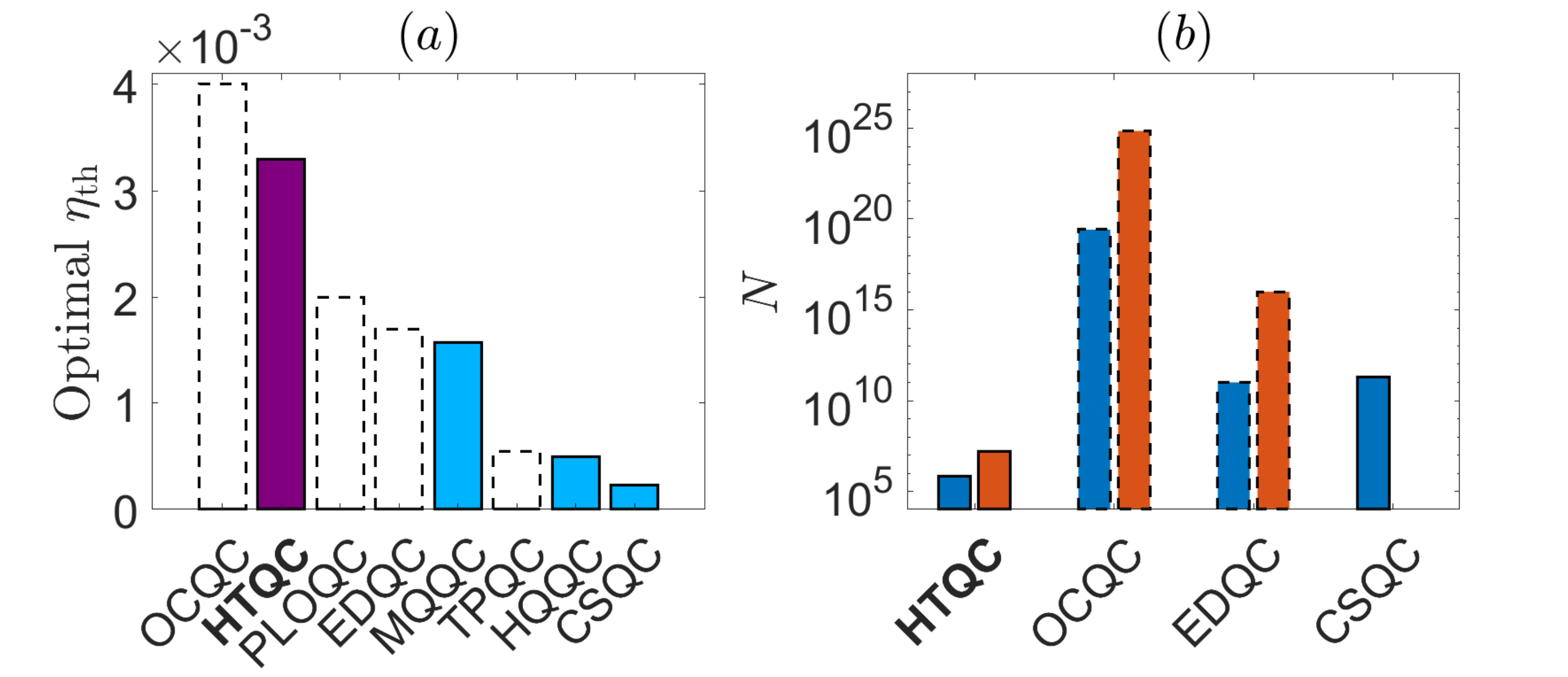}
\caption{(a) Optimal photon-loss threshold $\eta_{\rm th}$ for various QC schemes. 
It should be  noted that  $\eta_{\rm{th}}$'s of OCQC, PLOQC, EDQC and TPQC (dashed borders) are valid only for zero computational error, which is physically unachievable. Other schemes evaluate optimal $\eta_{\rm th}$ at nonzero computational errors  naturally related to $\eta$.
  (b) Resource overhead $N$ to achieve logical error rate  $p_{\rm L}\sim10^{-6}({\rm blue~shorter~ bars})$ and $p_{\rm L}\sim10^{-15}({\rm orange~taller~ bars})$ in terms of the average numbers of hybrid qubits (HTQC), entangled photon pairs (OCQC and EDQC), coherent-state superpositions (CSQC) from our analysis and published data in \cite{DHN06,Cho07,LRH08}. 
For CSQC data only for $p_{\rm L}\sim10^{-6}$ is available \cite{LRH08}. Obviously, HTQC is practically favorable for large scale QC both in terms of $\eta_{\rm th}$ and  $N$. 
See  Appendix~\ref{app:resource} for more details of comparisons. }
\label{fig:bar}
\end{figure}

Figure~\ref{fig:result}(b) shows the behavior of $\eta_{\rm th}$ with $\alpha$. 
Owing to the trade-off between $p_f$ and $p_Z$,  the optimal value for HTQC is $\alpha_{\rm opt}\approx1.25$ which corresponds to $\eta_{\rm th}\approx 3.3\times 10^{-3}$ and $p_{Z,\rm th}\approx6.9\times 10^{-3}$. The value of $\eta_{\rm th}$ for $0.8\leq\alpha\leq2$ is on the order of $10^{-3}$, which is an order greater than the non-topological hybrid-qubit-based QC (HQQC)~\cite{LJ13} and coherent state QC (CSQC)~\cite{LRH08}. HTQC also outperforms the  DV based topological photonic QC (TPQC)  with $\eta_{\rm th}\approx5.5\times10^{-4}$~\cite{HFJR10}. Multi-photon qubit QC (MQQC)~\cite{LPRJ15},  parity state linear optical QC (PLOQC)~\cite{ HHGR10} and error-detecting quantum state transfer based QC (EDQC) \cite{Cho07} provide $\eta_{\rm th}$'s which are  less than HTQC but of the same order as illustrated in Fig.~\ref{fig:bar}(a). In addition,  $\eta$ and the computational error rates are independent in~\cite{HHGR10, DHN06, Cho07}, while these two quantities are related  in our scheme and Refs.~\cite{LJ13,LRH08, LPRJ15}. 
Also in the former schemes the computational error is dephasing in nature, and in the latter schemes it is depolarizing.  In fact, $\eta_{\rm th}$'s claimed by optical cluster-state QC (OCQC)~\cite{DHN06},  PLOQC, EDQC and TPQC are valid  {\it only for  zero} computational error. This is unrealistic because photon losses typically cause computational errors.  For the computational error rate as low as $8\times 10^{-5}$,  $\eta_{\rm th}=0$ in OCQC. Thus, for  {\it non-zero} computational errors, HTQC also outperforms  OCQC due to its topological nature of QEC.

To estimate the resource overhead per gate operation, we count the average number of hybrid qubits $N$ required to build $\ket{\mathcal{C}_\mathcal{L}}$ of a sufficiently large side length $l$, where the desired value of $l$ depends on the target $p_{\rm L}$. The length $l$ is determined such that $\ket{\mathcal{C}_\mathcal{L}}$ can accommodate defects of circumference $d$ which are separated by distance $d$~\cite{WF14}. For this, the length of sides must be at least $l=5d/4$. Extrapolating the suppression of $p_{\rm L}$ with code distance, we determine the value of $d$ required to achieve the target $p_{\rm L}$ using the expression $p_{\rm L}= a^\prime/[(a/a^\prime)^{(d-d_{a^\prime})/2}]$~\cite{WF14}, where  $a$ and $a^\prime$ are  values of $p_{\rm L}$ corresponding to the second highest and the highest distances,  $d_{a}$ and $d_{a^\prime}$, chosen for simulation.   
Once  $d$ is determined, $N$ can be estimated as follows. Recall that two $\ket{\mathcal{C}_3}$'s and a $\ket{\mathcal{C}_{3^\prime}}$ are needed to build a $\ket{\mathcal{C}_\ast}$. On average, $8/[(1-\E{-2\alpha^{\prime2}})^2]$ hybrid qubits are needed to create a three-hybrid-qubit cluster ( as shown in Appendix~\ref{app:clus3}) and a total of  $24/[(1-\E{-2\alpha^{\prime 2}})^2]$ hybrid qubits for a $\ket{\mathcal{C}_\ast}$. Each $\ket{\mathcal{C}_\ast}$ corresponds to a single hybird qubit in the $\ket{\mathcal{C}_\mathcal{L}}$ and thus the number of $\ket{\mathcal{C}_\ast}$'s needed is $6l^3$. Finally, on average,   
$1125d^3/[4(1-\E{-2\alpha^{\prime 2}})^2]$ hybrid qubits are incurred.  For the optimal value of $\alpha_{\rm opt}\approx1.25$, from Fig.~\ref{fig:result}(a) we have $a\approx4.4\times10^{-4}$, 
$a^\prime\approx 7.9\times 10^{-5}$ and $d_{a^\prime}=9$; using these in the expression for $p_{\rm L}$ we find that $d \approx 14~(38)$ is needed to achieve $p_{\rm L}\sim10^{-6}~(10^{-15})$. This incurs  $N\approx8.5\times10^{5}~(1.7\times10^{7})$ hybrid qubits.

Comparisons in Fig.~4(b) and in Appendix~\ref{app:resource} show that HTQC incurs resources significantly less than all the other schemes under consideration.
As an example, for the case of TPQC,
we find $a=0.065$ and $a^\prime=0.059$ from Fig.~7(a) of \cite{HFJR10}, 
where the figure considers only computational errors.
Thus, TPQC under computational errors needs $d=225~(621)$ to attain $p_{\rm L}\sim10^{-6}~(10^{-15})$. Since a qubit in TPQC needs $2R+1$  photons on average as resources~\cite{HFJR10}, 
we obtain $N=(2R+1)\times6(5d/4)^3$ (derived in Appendix~\ref{app:resource}), where $R=7$ for maximum $\eta_{\rm th}$~\cite{HFJR10}. 
We then find $N=2\times10^{9}$~$(4.2\times10^{10})$ for TPQC,
and it must be even larger when qubit losses are considered together with computational errors  (see Appendix~\ref{app:resource} for an elaborate discussion).

\emph{Discussion.---}
Our proposal permits the construction of cluster states with very few missing edges that subsequently support QEC and QC  only with photon on-off measurements. We simulated its performance and found that our scheme is significantly more efficient  than other known schemes in terms of  both resource overheads and photon-loss thresholds (Fig.~\ref{fig:bar}), especially when exceedingly small logical error rates are desired for large-scale QC. We have considered measurements only on DV modes of hybrid qubits for QEC. However, measurements on  CV modes can also be used, which will significantly reduce leakage errors and improve the photon-loss threshold. The scheme requires hybrid qubits of $\alpha\approx\sqrt{2}\times1.25$  as raw resource states, which can  in principle be generated using 
available optical sources, linear optics and photodetectors \cite{JZK+14, MHL+14,KJ15}. 

One may examine other decoders tailored to take advantage of dephasing noise  instead of minimum weight perfect match~\cite{Fow12}, such as in~ \cite{TBF18}, for improvement of the photon-loss threshold. Different single-qubit noise models \cite{OSB13} may be considered to study the performance of HTQC. A  sideline task would be in-situ noise characterization using the available syndrome data~\cite{OSB15,OSB15s,OSB16,FSK+14}. The procedure proposed here to build complex hybrid clusters can also be used to build lattices of other geometries for QC~\cite{BM07, GSBR15, ZDLR} and other tasks such as communication~\cite{KKH15}. 

\begin{acknowledgments} 
We thank A.~G.~Fowler for useful discussions and S.-W.~Lee for providing data from~\cite{LJ13} used in Fig.~\ref{fig:result}.  
This work was supported by National Research Foundation of Korea (NRF) grants funded by the Korea government (Grants No.~2019M3E4A1080074 and No.~2020R1A2C1008609).
Y.S.T. was supported by an NRF grant funded by the Korea government  (Grant No. NRF-2019R1A6A1A10073437).
\end{acknowledgments} 

\appendix

\section{Bell State Measurement on hybrid qubits}
\label{app:bsm}
The hybrid-Bell state measurement (HBSM) depicted in Fig.~1(a) of the main Letter is composed of two types of measurements, $\Ba$  and $\rm B_s$, acting on the CV and DV modes, respectively. 
The $\Ba$ is successful with four possible outcomes that projects the input states of two hybrid qubits  onto one of the four possible hybrid-Bell sates
\cite{LJ13}: 
\bea
\ket{\psi^\pm}&=&\frac{1}{\sqrt{2}}(\ket{\alpha,\alpha}\ket{\textsc{h},\textsc{h}}\pm\ket{-\alpha,-\alpha}\ket{\textsc{v},\textsc{v}}),\nonumber\\
\ket{\phi^\pm}&=&\frac{1}{\sqrt{2}}(\ket{\alpha,-\alpha}\ket{\textsc{h},\textsc{v}}\pm\ket{-\alpha,\alpha}\ket{\textsc{v},\textsc{h}}).
\eea
Action of $\Ba$ results as clicks: (even, 0), (odd, 0), (0, even) and (0, odd) on the two  photon number parity detectors (PNPD), shown in Fig.~1~(a) of the primary manuscript, when projected on to $\ket{\psi^+}$, $\ket{\psi^-}$, $\ket{\phi^+}$ and $\ket{\phi^-}$, respectively.
To see the relation between the clicks on the PNPDs and the hybrid-Bell states,  pass the continuous variable (CV) modes  through the beam splitter (BS).  The states transform as $\ket{\psi^\pm}\rightarrow\frac{1}{\sqrt{2}}(\ket{\textsc{h},\textsc{h}}+\ket{\textsc{v},\textsc{v}})(|\sqrt{2}\alpha\rangle\pm|-\sqrt{2}\alpha\rangle)\ket{0}$, $\ket{\phi^\pm}\rightarrow\frac{1}{\sqrt{2}}(\ket{\textsc{h},\textsc{v}}+\ket{\textsc{v},\textsc{h}})\ket{0}(|\sqrt{2}\alpha\rangle\pm|-\sqrt{2}\alpha\rangle)$.  However, there is a possibility of having no clicks on both the PNPDs resulting in failure of the $\Ba$.  
The probability  of failure of $\Ba$ on hybrid qubits is $e^{-2\alpha^2}$. 

In spite of failure of the $\Ba$, it is still possible to carry out the Bell measurements using $\rm{B}_{\rm{s}}$ on the discrete variable (DV) modes of the hybrid qubits. The operation $\rm{B}_{\rm{s}}$ performs projection onto the states of 
$(\ket{\textsc{h},\textsc{h}}\pm\ket{\textsc{v},\textsc{v}})/\sqrt{2}$  and is successful with probability 1/2 only when both the photon detectors (PD) click together.  The HBSM on hybrid qubits fails only when both $\Ba$ and $\rm{B}_{\rm{s}}$ fail. Therefore, the probability of failure of HBSM is $\frac{1}{2}e^{-2\alpha^2}$ which rapidly approaches to zero as $\alpha$ grows.

\section{Generation of off-line resource states}
\label{app:clus3}
\begin{figure}[h]
\includegraphics[width=.8\columnwidth]{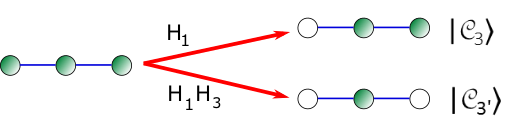}
\caption{Schematic diagram representing the transformation of 3-hybrid-qubit cluster state into $\ket{\mathcal{C}_3}$ and  $\ket{\mathcal{C}_{3^\prime}}$. ${\rm H}_1$ and ${\rm H}_3$ are the Hadamard operations on the first and the third (from the left) modes of the hybrid qubit, respectively. A unfilled circle represents the action of a  Hadamard transform. 
}
\label{fig:hadamard}
\end{figure}

Two kinds of three-hybrid-qubit cluster states,
\bea
\label{eq:offline}
\ket{\mathcal{C}_3}&=&
\frac{1}{2}\big(\ket{\alpha,\alpha,\alpha}  \ket{\textsc{h},\textsc{h},\textsc{h}}+ \ket{\alpha,\alpha,-\alpha}\ket{\textsc{h},\textsc{h},\textsc{v}} \nonumber\\
&&+\ket{-\alpha,-\alpha,\alpha}  \ket{\textsc{v},\textsc{v},\textsc{h}} - \ket{-\alpha,-\alpha,-\alpha}\ket{\textsc{v},\textsc{v},\textsc{v}}\big)\,,\nonumber\\
\ket{\mathcal{C}_{3^\prime}}&=&
\frac{1}{\sqrt{2}}\big(\ket{\alpha,\alpha,\alpha}  \ket{\textsc{h},\textsc{h},\textsc{h}}+ \ket{-\alpha,-\alpha,-\alpha}\ket{\textsc{v},\textsc{v},\textsc{v}}\big)
\eea
are used as offline resources to ballistically generate the {\it Raussendorf} lattice  $\ket{\mathcal{C}_\mathcal{L}}$. Equation~(\ref{eq:offline}) is an alternative expression of Eq.~(1) of the main manuscript where $\ket{0_{\rm L}}=\ket{\alpha}\ket{\textsc{h}}, ~\ket{1_{\rm L}}=\ket{-\alpha}\ket{\textsc{v}}$. These two states  $\ket{\mathcal{C}_3}$ and $\ket{\mathcal{C}_{3^\prime}}$ are the trsnsformation of the the linear 3-hybrid-qubit cluster state~\cite{BR01} as shown in Fig.~\ref{fig:hadamard} with all the hybrid-qubits filled.  The linear 3-hybrid-qubit cluster state~\cite{BR01}  has the form:
\begin{eqnarray}
\frac{1}{2\sqrt{2}}\big(\ket{0_{\rm L},0_{\rm L},0_{\rm L}}+\ket{0_{\rm L},0_{\rm L},1_{\rm L}}+\ket{0_{\rm L},1_{\rm L},0_{\rm L}}-\ket{0_{\rm L},1_{\rm L},1_{\rm L}} \nonumber \\
+\ket{1_{\rm L},0_{\rm L},0_{\rm L}}+ \ket{1_{\rm L},0_{\rm L},1_{\rm L}}-\ket{1_{\rm L},1_{\rm L},0_{\rm L}}+\ket{1_{\rm L},1_{\rm L},1_{\rm L}}\big).
\nonumber
\end{eqnarray}
We note that two kinds of off-line resource states are needed to generate larger cluster states {\it via} HBSMs because the Hadamard gate should be acted on one of the two input hybrid qubits~\cite{ZDLR}. Otherwise, the resulting states would be GHZ states rather than the desired cluster states. One can also verify this straightforwardly. It is important to note that the transformation shown in Fig.~\ref{fig:hadamard} or $\ket{\mathcal{C}_3}\leftrightarrow \ket{\mathcal{C}_{3^\prime}}$ is not possible via local operations on the hybrid qubits. To circumvent this issue, two types of qubit clusters  $\ket{\mathcal{C}_3}$ and $\ket{\mathcal{C}_{3^\prime}}$ need to be generated independently. This strategy is also efficient as creation of the linear 3-hybrid-qubit cluster states needs more hybrid qubits, $\Ba$s and $\BI$'s.

\begin{figure}[h]
\includegraphics[width=.7\columnwidth]{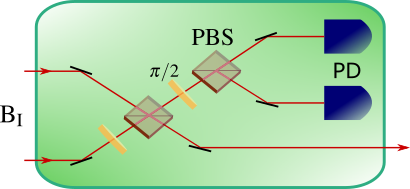}
\caption{$\BI$ is an entangling operation, which acts on the DV modes of  hybrid qubits and outputs one mode. Two PDs follow two polarizing beam splitter (PBS) with two $\pi/2$-rotators (wave plates) as shown in the figure.  $\BI$  is successful with probability $1/2$ only when one of the PDs clicks. 
}
\label{fig:b1}
\end{figure}

\begin{figure}[t]
\includegraphics[width=.95\columnwidth]{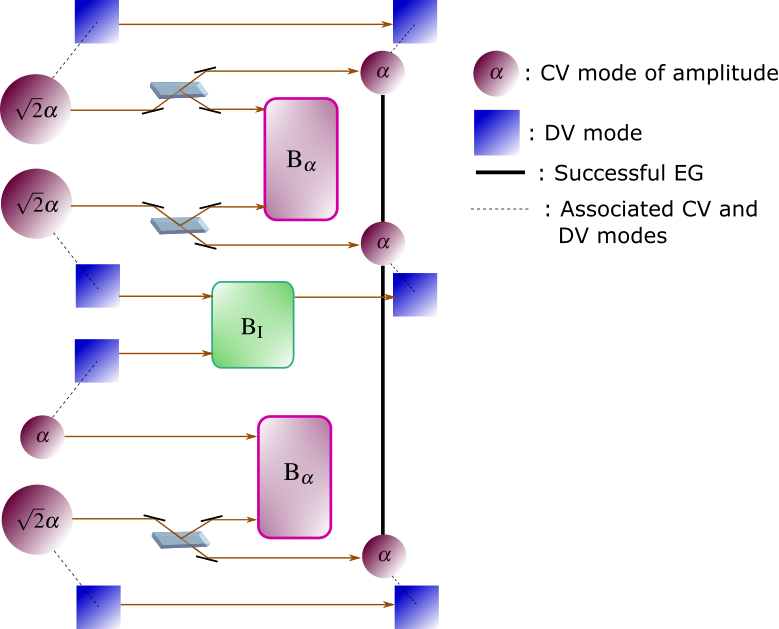}
\caption{
 A schematic diagram to build the offline resource state $\ket{\mathcal{C}_3}$ using  two $\Ba$'s, one $\BI$, three 1:1 beam splitters and four hybrid qubits, and EO represents entangling operation. To create  $\ket{\mathcal{C}_{3^\prime}}$, the ${\pi}/{2}$-rotator at the input of the $\BI$ is removed.
}
\label{fig:gates}
\end{figure}

The state $\ket{\mathcal{C}_3}$ is generated by entangling  three  $|+_{\rm L}^{\sqrt{2}\alpha}\rangle$'s (hybrid qubit with CV mode of amplitude $\sqrt{2}\alpha$)  and a  
$|+_{\rm L}\rangle$ using two $\Ba$'s and a $\BI$. As shown in Fig.~\ref{fig:b1}, $\BI$ has two polarizing BSs (PBS), $\frac{\pi}{2}$-rotators and and two photon detectors (PD). PBS transmits $\ket{\textsc{v}}$ and reflects $\ket{\textsc{h}}$.   $\BI$ performs the operation $\ket{\textsc{h}}\langle \textsc{h},\textsc{h}|+\ket{\textsc{h}}\langle \textsc{h},\textsc{v}|+\ket{\textsc{v}}\langle \textsc{v},\textsc{h}|-\ket{\textsc{v}}\langle \textsc{v},\textsc{v}|$ \bla which succeeds with probability 1/2 only when one of the PDs click. Instances with no click and both PDs clicking are failures.    Hybrid qubits are initialized as $|+_{\rm L}^{\sqrt{2}\alpha}\rangle\otimes|+_{\rm L}^{\sqrt{2}\alpha}\rangle\otimes|+_{\rm L}\rangle\otimes|+_{\rm L}^{\sqrt{2}\alpha}\rangle$  
and passed on to beam splitters  as shown in Fig.~\ref{fig:gates}. The resulting state of the hybrid qubits is

\bea
\label{eq:4sep}
&& \ket{\alpha,\alpha,\alpha,\alpha,\alpha,\alpha,\alpha} \ket{\textsc{h},\textsc{h},\textsc{h},\textsc{h}}\nonumber\\
&&+ \ket{\alpha,\alpha,\alpha,\alpha,\alpha,-\alpha,-\alpha} \ket{\textsc{h},\textsc{h},\textsc{h},\textsc{v}}\nonumber\\
&& +\ket{\alpha,\alpha,\alpha,\alpha,-\alpha,\alpha,\alpha} \ket{\textsc{h},\textsc{h},\textsc{v},\textsc{h}}\nonumber\\
&&+\ket{\alpha,\alpha,\alpha,\alpha,-\alpha,-\alpha,-\alpha} \ket{\textsc{h},\textsc{h},\textsc{v},\textsc{v}}\nonumber\\
&& +\ket{\alpha,\alpha,-\alpha,-\alpha,\alpha,\alpha,\alpha} \ket{\textsc{h},\textsc{v},\textsc{h},\textsc{h}}\nonumber\\
&&+ \ket{\alpha,\alpha,-\alpha,-\alpha,\alpha,-\alpha,-\alpha} \ket{\textsc{h},\textsc{v},\textsc{h},\textsc{v}}\nonumber\\
 &&+\ket{\alpha,\alpha,-\alpha,-\alpha,-\alpha,\alpha,\alpha} \ket{\textsc{h},\textsc{v},\textsc{v},\textsc{h}}\nonumber\\
&&+\ket{\alpha,\alpha,-\alpha,-\alpha,-\alpha,-\alpha,-\alpha} \ket{\textsc{h},\textsc{v},\textsc{v},\textsc{v}}\nonumber\\
&& +\ket{-\alpha,-\alpha,\alpha,\alpha,\alpha,\alpha,\alpha} \ket{\textsc{v},\textsc{h},\textsc{h},\textsc{h}}\nonumber\\
&& +\ket{-\alpha,-\alpha,\alpha,\alpha,\alpha,-\alpha,-\alpha} \ket{\textsc{v},\textsc{h},\textsc{h},\textsc{v}}\nonumber\\
&& +\ket{-\alpha,-\alpha,\alpha,\alpha,-\alpha,\alpha,\alpha} \ket{\textsc{v},\textsc{h},\textsc{v},\textsc{h}}\nonumber\\
&&+\ket{-\alpha,-\alpha,\alpha,\alpha,-\alpha,-\alpha,-\alpha} \ket{\textsc{v},\textsc{h},\textsc{v},\textsc{v}}\nonumber\\
&& +\ket{-\alpha,-\alpha,-\alpha,-\alpha,\alpha,\alpha,\alpha} \ket{\textsc{v},\textsc{v},\textsc{h},\textsc{h}}\nonumber\\
&& +\ket{-\alpha,-\alpha,-\alpha,-\alpha,\alpha,-\alpha,-\alpha} \ket{\textsc{v},\textsc{v},\textsc{h},\textsc{v}}\nonumber\\
&& +\ket{-\alpha,-\alpha,-\alpha,-\alpha,-\alpha,\alpha,\alpha} \ket{\textsc{v},\textsc{v},\textsc{v},\textsc{h}}\nonumber\\
&&+\ket{-\alpha,-\alpha,-\alpha,-\alpha,-\alpha,-\alpha,-\alpha} \ket{\textsc{v},\textsc{v},\textsc{v},\textsc{v}}.
\eea

Upon successful $\Ba(2,3)$ and  $\Ba(5,6)$  with respective outcomes, say  $\ket{\psi^+}$ and $\ket{\psi^+}$, the state in  Eq.~(\ref{eq:4sep}) reduces to
 \bea
&& \ket{\alpha,\alpha,\alpha}  \ket{\textsc{h},\textsc{h},\textsc{h},\textsc{h}}+ \ket{\alpha,\alpha,-\alpha}\ket{\textsc{h},\textsc{h},\textsc{v}, \textsc{v}}+ \nonumber\\
 &&\ket{-\alpha,-\alpha,\alpha}  \ket{\textsc{v},\textsc{v},\textsc{h},\textsc{h}} + \ket{-\alpha,-\alpha,-\alpha}\ket{\textsc{v},\textsc{v},\textsc{v},\textsc{v}},\nonumber
 \eea  
 where $\Ba(n,m)$ represents the action of $\Ba$ on $n$-th and $m$-th CV modes as shown in the Fig.~\ref{fig:gates}.
Further, with the successful  $\BI(2,3)$ (DV modes 2 and 3 being the inputs as shown in the Fig.~\ref{fig:gates}) we get 
 \bea
\label{eq:clus3}
&& \ket{\mathcal{C}_3}=\ket{\alpha,\alpha,\alpha}  \ket{\textsc{h},\textsc{h},\textsc{h}}+ \ket{\alpha,\alpha,-\alpha}\ket{\textsc{h},\textsc{h},\textsc{v}} +\nonumber\\
 &&\ket{-\alpha,-\alpha,\alpha}  \ket{\textsc{v},\textsc{v},\textsc{h}} - \ket{-\alpha,-\alpha,-\alpha}\ket{\textsc{v},\textsc{v},\textsc{v}}.
\eea 
Note that for other possible measurement outcomes on $\Ba$ and $\BI$, the  $\ket{\mathcal{C}_3}$ will be equivalent up to local Pauli operations. The local operations to be performed upon getting different measurement comes are listed in Table~\ref{tab:clus3}.
The logical Pauli operations on hybrid qubits can be accomplished with the polarization rotator on the DV mode and the $\pi$-phase shifter on the CV mode. $X_\textrm{L}$: $\ket{\alpha}\leftrightarrow\ket{-\alpha},~ \ket{\textsc{h}}\leftrightarrow\ket{\textsc{v}}$ and $Z_\textrm{L}$:  $\ket{\textsc{h}}\rightarrow\ket{\textsc{h}}$, $ \ket{\textsc{v}}\rightarrow-\ket{\textsc{v}}$ with no need for  action on $\ket{\alpha}$. These local operations are used only in creating the offline resource states, which is not a ballistic process. 

Similarly, state  $\ket{\mathcal{C}_{3^\prime}}$ can be generated  by removing the $\pi/2$-rotator at the input of the $\BI$ in Fig.~\ref{fig:b1}.  Here, the only other possible outcome on the $\Ba$s is 
$\ket{\psi^-}$, in which case  the relative phase can be changed by applying a $Z_{\rm L}$ for the measurement outcome combination $\{\ket{\psi^\pm},\ket{\psi^\mp}\}$. 
The kets $\ket{\mathcal{C}_3}$ and $\ket{\mathcal{C}_{3^\prime}}$ are created only when the operations $\Ba$ and $\BI$ succeed together. The probability that all the three operations are successful is  $\frac{1}{2}(1-e^{-2\alpha^2})^2$. Thus, the average number of hybrid qubits consumed for generating a  $\ket{\mathcal{C}_3}$ or $\ket{\mathcal{C}_{3^\prime}}$  is $8/(1-e^{-2\alpha^2})^2$.

\begin{table}[b]
\begin{tabular}{|c|c|c|c|}\hline
$ \rm B_{\alpha}(2,3)$ & $\rm B_{\alpha}(5,6)$ & $\BI(2,3)$ & Local Operation   \\ \hline
$ \ket{\psi^+}$&$ \ket{\psi^+}$  &\textsc{h}/\textsc{v} & NA/$Z_3$  \\
$ \ket{\psi^+}$ & $ \ket{\psi^-}$ &   & $Z_3$/$Z_2Z_3$ \\
$ \ket{\psi^+}$ & $ \ket{\phi^+}$ &    &$Z_2$/NA\\ 
$ \ket{\psi^+}$ & $ \ket{\phi^-}$ &    & $Z_2Z_3$/$X_3Z_2Z_3$\\ \hline
$ \ket{\psi^-}$&$ \ket{\psi^+}$  &\textsc{h}/\textsc{v} & $Z_2$/NA  \\
$ \ket{\psi^-}$ & $ \ket{\psi^-}$ &   &$Z_2Z_3$/$Z_3$ \\
$ \ket{\psi^-}$ & $ \ket{\phi^+}$ &    & NA/$Z_2$\\ 
$ \ket{\psi^-}$ & $ \ket{\phi^-}$ &    & $X_3Z_2Z_3$/$Z_2Z_3$\\ \hline
$ \ket{\phi^+}$&$ \ket{\psi^+}$  &\textsc{h}/\textsc{v} & $X_1$/$X_1Z_1$   \\
$ \ket{\phi^+}$ & $ \ket{\psi^-}$ &   &$X_2$/$X_1Z_2Z_3$ \\
$ \ket{\phi^+}$ & $ \ket{\phi^+}$ &    &$ X_1Z_2$/$X_1$\\ 
$ \ket{\phi^+}$ & $ \ket{\phi^-}$ &    & $X_1Z_2Z_3$/$X_1Z_3$\\ \hline
$ \ket{\phi^-}$&$ \ket{\psi^+}$  &\textsc{h}/\textsc{v} & $X_2Z_2Z_3$/$X_1$  \\
$ \ket{\phi^-}$ & $ \ket{\psi^-}$ &   & $X_2Z_2$/$X_2$\\ 
$ \ket{\phi^-}$ & $ \ket{\phi^+}$ &    &$X_1$/$X_1Z_2$\\ 
$ \ket{\phi^-}$ & $ \ket{\phi^-}$ &    &$X_2$/$X_1Z_2Z_3$\\ \hline
\end{tabular}
\caption{ The table lists the local operations to be performed upon getting different measurement comes on $\Ba$ and $\BI$}
\label{tab:clus3}
\end{table}

\section{Hybrid qubits under photon loss}
\label{app:loss}
The action of the photon-loss channel $\mathcal{E}$ on a hybrid qubit initialized in the state  $\rho_0=\ket{+_{\rm L}}\langle +_{\rm L}|$ is \cite{LJ13}
\bea
\label{eq:loss}
&\mathcal{E}(\rho_0)&=(1-\eta)\bigg(\frac{1+e^{-2\eta\alpha^2}}{2}\ket{+^\prime}\langle+^\prime|+\nonumber\\
&&+\frac{1-e^{-2\eta\alpha^2}}{2}\ket{-^\prime}\langle -^\prime|\bigg)
+\frac{\eta}{2}\left(|+^{\rm lk}\rangle\langle +^{\rm lk}|+|-^{\rm lk}\rangle\langle -^{\rm lk}|\right)\nonumber\\
&=&\frac{(1-\eta)}{2}\bigg(|\alpha^\prime,\textsc{h}\rangle\langle\alpha^\prime,\textsc{h}|+|-\alpha^\prime,\textsc{v}\rangle\langle-\alpha^\prime,\textsc{v}|\nonumber\\
&+&e^{-2\eta\alpha^2}\big(|\alpha^\prime,\textsc{h}\rangle\langle-\alpha^\prime,\textsc{v}|+|-\alpha^\prime,\textsc{v}\rangle\langle\alpha^\prime,\textsc{h}|\big)\bigg)\nonumber\\
&+&\frac{\eta}{2}\bigg(\big(|\alpha^\prime\rangle\langle\alpha^\prime|+|-\alpha^\prime\rangle\langle-\alpha^\prime|\big)\otimes\ket{0}\langle0|\bigg)
\eea
where $\ket{\pm^\prime}=(\ket{\alpha^\prime,\textsc{h}}\pm\ket{-\alpha^\prime,\textsc{v}})/\sqrt{2},~ \ket{\pm^{\rm lk}}=(\ket{\alpha^\prime}\pm\ket{-\alpha^\prime})\otimes\ket{0}/\sqrt{2} $ and $\alpha^\prime=\sqrt{1-\eta}\alpha$ with $\eta$ being photon loss rate which also models imperfect sources, detectors, or absorptive optical components.
It can be seen from the Eq.~(\ref{eq:loss}), that due to photon loss the CV part is dephased, i.e., a phase flip error $Z$ occurs on the hybrid qubits and the loss on the DV part makes them leak out of the logical basis.
Also, due to photon losses the success rate of the $\Ba$ reduces to $(1-e^{-2\alpha^{\prime2}})$ and that of $\BI$ remains same. As a result, the average number of hybrid qubits  to build off-line resource states increases to $8/[(1-e^{-2\alpha^{\prime2}})^2]$.
The failure rate of the HBSM, $p_f$ increases to $(1-\eta)\E{-2\alpha^{\prime 2}}/2+\eta\E{-2\alpha^{\prime 2}}= (1+\eta)\E{-2\alpha^{\prime 2}}/2$, where $\alpha^\prime=\sqrt{1-\eta}\alpha$.  The first term originates from the attenuation of CV mode, and the second from both CV attenuation and DV loss.

{\it Loss-tolerance of  $\BI$}:
One can verify that the noisy DV part in Eq.~(\ref{eq:clus3}), $\mathcal{E}^{\otimes4}( \ket{\textsc{h},\textsc{h},\textsc{h},\textsc{h}}+ \ket{\textsc{h},\textsc{h},\textsc{v}, \textsc{v}} 
 +  \ket{\textsc{v},\textsc{v},\textsc{h},\textsc{h}} + \ket{\textsc{v},\textsc{v},\textsc{v},\textsc{v}})$ is transformed in to $\mathcal{E}^{\otimes3}( \ket{\textsc{h},\textsc{h},\textsc{h}}+ \ket{\textsc{h},\textsc{h}, \textsc{v}}  +  \ket{\textsc{v},\textsc{v},\textsc{h}} -\ket{\textsc{v},\textsc{v},\textsc{v}})$ under the action of $\BI$. The noise channel on the resulting state is still
 $\mathcal{E}$, implying that no additional computational errors  are introduced by  $\BI$, unlike in Ref.~\cite{LHMB15}.

\subsection{Noise by HBSM under photo-loss}
\label{app:ba}
\begin{figure}[h]
\includegraphics[width=.7\columnwidth]{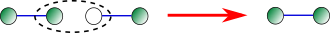}
\caption{Typical instance of action of HBSM on the noisy hybrid cluster states.
 }
\label{fig:clus2}
\end{figure}
Let us consider the typical situation where HBSMs are used to create entanglement between the desired hybrid qubits as shown in the Fig.~\ref{fig:clus2}. Looking at only the CV part, we have
\be
\label{eq:ba}
(\ket{\alpha,\alpha}+\ket{\alpha,-\alpha}+\ket{-\alpha,\alpha}-\ket{-\alpha,-\alpha})(\ket{\alpha,\alpha}+(\ket{-\alpha,-\alpha}).
\ee 
To determine the noise on the resultant cluster state after the action of $\Ba$, we consider the hybrid qubits undergoing BSM to be noisy. The resulting noisy state, in the logical basis is of the form 
\be
\label{bs1}
\frac{1}{4}\begin{pmatrix}
1 &e^{-4\eta\alpha^2}&1&e^{-4\eta\alpha^2}\\
e^{-4\eta\alpha^2}&1&e^{-4\eta\alpha^2}&1\\
1&e^{-4\eta\alpha^2}&1&e^{-4\eta\alpha^2}\\
e^{-4\eta\alpha^2}&1&e^{-4\eta\alpha^2}&1
\end{pmatrix}.
\ee
For the desired result of the HBSMs, the Hadamard could be on the other hybrid qubit too where the terms in the Eq.~\ref{eq:ba} are interchanged and the resulting noisy sate is 
\be
\label{bs2}
\frac{1}{4}\begin{pmatrix}
1 &1&e^{-4\eta\alpha^2}&e^{-4\eta\alpha^2}\\
1 &1&e^{-4\eta\alpha^2}&e^{-4\eta\alpha^2}\\
e^{-4\eta\alpha^2}&e^{-4\eta\alpha^2}&1&1\\
e^{-4\eta\alpha^2}&e^{-4\eta\alpha^2}&1&1\\
\end{pmatrix}.
\ee
Considering both the instances of action of $\Ba$, the resultant state is the equal weighted superposition of the states in Eqs.~(\ref{bs1}) and (\ref{bs2}), and the corresponding Kraus operators for this noisy channel $\mathcal{E}^{\Ba}$ are
$\Big\{ \sqrt{\frac{1+e^{-4  \eta\alpha^{2}}}{2}}  I\otimes I,~~ \sqrt{\frac{1-e^{-4  \eta\alpha^{2}}}{4}}  Z\otimes I,~~ \sqrt{\frac{1-e^{-4  \eta\alpha^{2}}}{4}}  I\otimes Z\Big\}$.
This noise channel $\mathcal{E}^{\Ba}$ is used as the noise due to entangling operation in the AUTOTUNE~\cite{FWMR12}. 
We observe that for $\eta\sim10^{-3}$ and small values of $\alpha$,  $p_Z\approx(1-e^{-4  \eta\alpha^2})/4$. Thus, the $\mathcal{E}^{\Ba}$ will have the following Kraus operators 
$\{ \sqrt{(1-2p_Z)}~ I\otimes I,~~\sqrt{p_Z}~  Z\otimes I,~~ \sqrt{p_Z} ~ I\otimes Z \}$. 

{\it Loss-tolerance of $\rm B_s$}:
Now, let us move our attention towards the DV part of the noisy cluster states as shown in the Fig.~\ref{fig:clus2} and study the action of ${\rm B_s}$ on them during photon loss. One can verify that the state
$\mathcal{E}\big(  \ket{\textsc{h},\textsc{h}}+ \ket{\textsc{v},\textsc{v}}\big)\otimes \big(\ket{\textsc{h},\textsc{h}} +\ket{\textsc{h},\textsc{v}}+\ket{\textsc{v},\textsc{h}}- \ket{\textsc{v},\textsc{v}}\big)$ is transformed to $\mathcal{E}\big(|\textsc{h},\textsc{h}\rangle+|\textsc{h},\textsc{v}\rangle+|\textsc{v},\textsc{h}\rangle-|\textsc{v},\textsc{v}\rangle\big)$ under the action of ${\rm B_s}$.
The noise channel on the resulting state is still $\mathcal{E}$ which confirms that the ${\rm B_s}$ introduces no additional computational errors.

\section{Short note on simulation with AUTOTUNE}
\label{app:sim}
To simulate topological quantum error correction (QEC) on a noisy $|\mathcal{C}_\mathcal{L}\rangle$, we use the software package  AUTOTUNE~\cite{FWMR12} which offers a wide range of options of noise models and the ability to customize them. Remarkably, it allows for simulation  of QEC when some qubits on  $|\mathcal{C}_\mathcal{L}\rangle$ are missing. This feature allows us to simulate missing edges by mapping them to missing qubits as explained in the main Letter. AUTOTUNE  uses the circuit model to simulate the error propagation while building $|\mathcal{C}_\mathcal{L}\rangle$. 

 A noisy $|\mathcal{C}_\mathcal{L}\rangle$ can be simulated using AUTOTUNE in the following way. We notice that in order to generate  $|\mathcal{C}_\mathcal{L}\rangle$, the action of HBSMs for creating  edges between the central hybrid qubits of $|\mathcal{C}_\ast\rangle$ (see Fig.~1(b) of the main Letter) is the equivalent to the action of controlled-Z (CZ) gates
on qubits all initialized in $\ket{+}$'s (eigenstates of Pauli-X operator with eigenvalue +1) on the lattice
   in the circuit model. So, the action of HBSMs under noise can be simulated by noisy CZ gates in AUTOTUNE and thereafter the whole QEC.  Also,  AUTOTUNE allows for the simulation of the noise added during the initialization of the qubits, noisy gate operations, and error propagation through the lattice.

AUTOTUNE used for simulation is optimized compared to the one used in Ref.~\cite{HFJR10}. In the decoding process, AUTOTUNE uses additional diagonal links compared to {\it Manhattan lattice} \cite{FWMR12}. These diagonal links avoid the wrong identification of error chains by the minimum-weight perfect-match algorithm and guarantee corrections of  $(d-1)/2$ errors to improve the accuracy in estimating the logical error rate $p_{\rm L}$. These details are in Sec.~IV  of Ref.~\cite{FWMR12}.
Thus the results obtained from the simulations of the topological QEC corresponding to  code distances $d=5,7,9$ are reliable and the finite-size effect is negligible. An example for this is Ref.~\cite{WF14} which uses AUTOTUNE to extrapolates the behaviors of higher distances from that of $d=5,7$ and sometimes $d=9$.


In HTQC, mapping the missing edges (due to failed HBSMs) to missing qubits is a crucial part in the simulation of QEC on a noisy $|\mathcal{C}_\mathcal{L}\rangle$. The HBSMs are used in two cases: (a) building star cluster $\ket{\mathcal{C}_\ast}$ and (b) connecting the  $\ket{\mathcal{C}_\ast}$'s as shown in  Fig.~2(b) of the main Letter. In the first case, failure of an HBSM creates a diagonal edge that must be removed by measuring the two qubits in the Z-basis, $M_z$, as shown in Fig.~2(c) of the main Letter. Thus, in the first case of failure of an HBSM, a missing edge corresponds to two missing qubits. 
In the second case, failure of an HBSM leads to a missing edge between the qubits  as shown in  Fig.~2(a) of the main Letter. During QEC one of the qubits associated with the missing edge is removed by $M_z$. Thus, in the second case of failure, a missing edge corresponds to one missing qubit.

In a lattice $|\mathcal{C}_\mathcal{L}\rangle$ of distance $d$,  one can count that there are $6d^3$ qubits  connected by $12d^3$ edges as done in Sec.~VIII of Ref.~\cite{WF14}. In this approach, a lattice of distance $d$, there are $d^3$ unit cells, and  each unit cell has effectively 6 qubits \cite{WF14}. Each face of the unit cell has four edges shared between two faces. Thus each face has effectively two edges. Thus a unit cell has on average 12 edges.

A star cluster state $\ket{\mathcal{C}_\ast}$ corresponds to one lattice qubit, and two HBSMs are used for building the star cluster state. Thus, an equal number ({\it i.e.,} $12d^3$ per lattice state $|\mathcal{C}_\mathcal{L}\rangle$) of HBSMs are used for  creation of $|\mathcal{C}_\ast\rangle$'s and for connecting them in HTQC. Considering both the cases of usage of HBSMs, on average 1.5 qubits are lost per HBSM failure.

As pointed out in the result section of the main Letter, the target error rate is estimated using the suppression ratio of $p_{\rm L}$ with $d$. As  AUTOTUNE uses the minimum-weight perfect-match algorithm, the suppression ratio is expected to be nearly constant when it is sufficiently away from the threshold and $d$ is large~\cite{WF14, Fow12}. If  the  suppression ratio is not constant, resource estimation will lead to an underestimation. From the simulation results, we have the suppression ratio between distances 3 and 5 to be $a/a^\prime\approx12.5$,  between 5 and 7 to be $a/a^\prime\approx6.1$, and between 7 and 9 to be $a/a^\prime\approx5.6$. We observe that the suppression ratio is nearly constant between distances 5 and 7 (7 and 9). Thus we choose the  suppression ratio between the distances 7 and 9, i.e., $a/a^\prime\approx5.6$ to estimate target $p_{\rm L}$ and resource overheads.

We observe from Fig.~3(a) of the main Letter that if we add curves of higher values of $d$, the threshold point $p_{Z,\rm th}$ would tend to shift towards the higher side. The same can be observed from Fig.~10 of  Ref.~\cite{RHG07}. However, taking a conservative approach we consider the value of $p_{Z,{\rm th}}$ to be $6.9\times10^{-3}$, which corresponds to photon-loss threshold of $\eta_{\rm th}=3.3\times10^{-3}$.

\section{Resource efficiency of HTQC against unreported schemes}
\label{app:resource}
There are no  reports estimating resource $N$ to attain a logical error rate  $p_{\rm L}$ of $10^{-6}$ or $10^{-15}$  for PLOQC, MQQC, HQQC and TPQC. The first three schemes are based on the 7-qubit Steane code which has a typical value of $p_{\rm L}\sim10^{-3}$ for the first level of telecorrection and needs more levels of concatenation  to reach a smaller value of $p_{\rm L}$. Typically, 4 levels suffice to attain $p_{\rm L}\sim10^{-6}$~\cite{DHN06,LRH08}.
In what follows, we make an informed guess to justify the resource efficiency of HTQC. 

\begin{enumerate}
\item 
The first level of telecorrection in HQQC needs approximately $5.35\times10^3$ hybrid-qubits (with $\alpha=1.1$)~\cite{LJ13}. 
The number of hybrid-qubits incurred per operation, at a particular level L, is
given by the number of hybrid-qubits consumed per error correction step at first level, multiplied by number of operations needed at levels $2, \dots ,L$~\cite{DHN06}. 
Typically, error correction at first level requires 1000 operations~[10]. Here, we assume that level 2,3 would require 100 operations and make resource estimation similar to that in Ref.~\cite{LJ13} for first level. 
Roughly speaking, adding another 3 levels would cost  $N\approx8\times10^9$ which shows that the HTQC is more resource-efficient.

\item 
MOQC requires the following multi-mode optical resource states (upto normalization)  for optimal loss-threshold~\cite{LPRJ15}:
$$\ket{Z}\equiv\ket{+}^{\otimes4}\ket{+}^{\otimes4}+\ket{+}^{\otimes4}\ket{-}^{\otimes4}+\ket{-}^{\otimes4}\ket{+}^{\otimes4}-\ket{-}^{\otimes4}\ket{-}^{\otimes4}$$
$$\ket{Z^\prime}\equiv\ket{+}^{\otimes8}\ket{+}^{\otimes8}+\ket{+}^{\otimes8}\ket{-}^{\otimes8}+\ket{-}^{\otimes8}\ket{+}^{\otimes8}-\ket{-}^{\otimes8}\ket{-}^{\otimes8},$$
 where $\ket{\pm}=(\ket{\textsc{h}}\pm\ket{\textsc{v}})/\sqrt{2}$. Using ${\rm B_I}$ of success rate 0.5 and ${\rm B_{s}}$ of boosted success rate 0.75 on the supply of Bell states, $\ket{Z}$ and 
$\ket{Z^\prime}$ can be constructed. Let's denote the $n$-mode GHZ state by $\ket{{\rm GHZ}_n}=\ket{\textsc{h}}^{\otimes n}+\ket{\textsc{v}}^{\otimes n}$ (upto normalization).

To generate $\ket{Z}$, we  fuse  $\ket{{\rm GHZ}_6}$ and  $\ket{{\rm GHZ}_4}$  using ${\rm B_{s}}$ with a Hadamard operation on the first mode of the latter state. The resulting state would be equivqlent to the $\ket{Z}$ upto local Hadamard operations. In prior, the $\ket{{\rm GHZ}_4}$ is generated with two $\ket{{\rm GHZ}_3}$ using a ${\rm B_{s}}$'s. Thus, on average $2/0.75$ number of $\ket{{\rm GHZ}_3}$ are needed. 
On the other hand, $\ket{{\rm GHZ}_6}$ is created by fusing two $\ket{{\rm GHZ}_4}$'s using a ${\rm B_{s}}$'s, which requires, on average, $(2/0.75)^2$ number of $\ket{{\rm GHZ}_3}$'s. 
Finally, $\ket{Z}$ needs, on average, $\frac{1}{0.75}\big(\frac{2}{0.75^2}+\frac{2}{0.75}\big)$ number of $\ket{{\rm GHZ}_3}$'s. Each $\ket{{\rm GHZ}_3}$ can be generated using ${\rm B_I}$ and needs four Bell states, on average. Additionally,  8 Bell states are spent for boosting the success rate of $\rm B_{s}$~\cite{G11}. Totally,  $\frac{1}{0.75}\big((\frac{2}{0.75})^2+\frac{2}{0.75}\big)\times4+8\approx 60$ Bell states are consumed to generate $\ket{Z}$. 

To generate $\ket{Z^\prime}$, one needs to fuse $\ket{{\rm GHZ}_{10}}$ and $\ket{{\rm GHZ}_8}$  using ${\rm B_{s}}$ with a Hadamard operation on the first mode of the latter state. On average,  $(2/0.75)^3$ number of $\ket{{\rm GHZ}_3}$'s are needed to generate a $\ket{{\rm GHZ}_{10}}$. Following the calculation carried out in the  case of generating $\ket{Z}$, finally,  $\frac{1}{0.75}\big[(\frac{2}{0.75})^3+\frac{1}{0.75}\big((\frac{2}{0.75})^2+\frac{2}{0.75}\big)\big]\times4+16\approx 187$ number of Bell states are incurred to build $\ket{Z^\prime}$.

 Following the procedure to calculate the resource cost as per Ref.~\cite{LJ13},
first level of error correction needs $7.2\times10^4$ Bell states. Arguing similar to the case of HQQC,  adding another 3 levels would cost $N\approx2.7\times10^{14}$. Thus, HTQC is better than MQQC in terms of resource efficiency. 

\item For PLOQC, $N\approx1.8\times10^5$ just for the first level  of telecorrection~\cite{HHGR10}, and making argument similar to HQQC, resource incurred for 4 levels  would be larger by many orders of magnitude. Hence, HTQC is more resource-efficient.

\item
By extrapolating the suppression of $p_{\rm L}$ with $d$, we estimate  $d$ required to achieve the target $p_{\rm L}\approx 10^{-6}(10^{-15})$ using the expression~\cite{WF14},
\be
p_{\rm L}=\frac{a^\prime}{(\frac{a}{a^\prime})^{(d-d_{a^\prime})/2}},
\label{eq:dis}
\ee
where $a$ and $a^\prime$ are the values of $p_{\rm L}$ corresponding to the second highest distance, $d_a$, and the highest distance, $d_{a^\prime}$, chosen for the simulation.
For the topological photonic QC (TPQC), we can make resource estimation by looking at Figs.~7(a) and (b) of Ref.~\cite{HFJR10}. 

First, we look into Fig.~7(a) of Ref.~\cite{HFJR10} that plots $p_{\rm L}$ against computational error rate and corresponds to the case with {\it no photon loss} (qubit loss). Here, 
$a\approx 0.065$ (corresponding to $d_a=13$) and $a^\prime\approx 0.059$ (corresponding to $d_{a^\prime}=15$).  Using Eq.~(\ref{eq:dis}) above, we find that a $\ket{\mathcal{C}_\mathcal{L}}$ of $d\approx 225~(621)$ is essential to attain $p_{\rm L}\sim10^{-6}~(10^{-15})$. 
In TPQC, {\it redundantly encoded photons} are used as a qubit in the $\ket{\mathcal{C}_\mathcal{L}}$ and to maximize the probability of the edge creation between the qubits. To create an edge successfully, the entangling operation is performed $R$ times. On average, this requires a qubit to consist of $2R+1$ photons~\cite{HFJR10}. We know that a $\ket{\mathcal{C}_\mathcal{L}}$ of side $5d/4$ would have $6\times(5d/4)^3$ qubits, on average. In TPQC the redundant encoded photons are considered as resources. Thus, the incurred resource overhead in TPQC would be 
$N=(2R+1)\times 6\big(\frac{5d}{4}\big)^3$.  For optimal $\eta_{\rm th}$, $R$ is set to 7 in TPQC~\cite{HFJR10}, and we obtain  $N\approx 2\times10^9~(4.2\times10^{10})$ for $d= 225~(621)$.

Next, we consider  Fig.~7(b) of Ref.~\cite{HFJR10}  where $p_{\rm L}$ is plotted against photon loss (qubit loss). We emphasize that in this plot there is {\it no computational error} introduced. Here, we have $a\approx 0.015$ (corresponding to $d_a=13$) and $a^\prime\approx 0.01$ (corresponding to $d_{a^\prime}=15$).  Using Eq.~(\ref{eq:dis}), it is found that $\ket{\mathcal{C}_\mathcal{L}}$ of $d\approx 60~(162)$ is essential to attain $p_{\rm L}\sim10^{-6}~(10^{-15})$.  In this case of TPQC  we get  $N\approx 3.8\times10^7~(7.5\times10^8)$.

Both the computational errors and the photon losses are considered {\it together} for estimating resource overheads of all the other schemes. If both the factors are considered for TPQC, the incurred resources would be  much more than $2\times10^9~(4.2\times10^{10})$ to attain $p_{\rm L}\sim10^{-6}~(10^{-15})$. This means that HTQC offers a resource efficiency better than that of TPQC at least by 3 orders of magnitude.
We note that considering only the  factor of photon loss  would lead to an underestimation of the resource overhead for TPQC.

\end{enumerate}

\bibliographystyle{apsrev4-1}

\end{document}